\documentclass[11pt]{article}

\usepackage[slantedGreek]{mathpazo}
\usepackage[pdftex]{graphicx}
\usepackage{amsfonts}
\usepackage{setspace}
\usepackage{amssymb}
\usepackage{amsthm}
\usepackage{graphicx}
\usepackage{amsmath}
\usepackage{mathtools}
\usepackage{lscape}
\usepackage{subcaption}
\usepackage{suffix}
\usepackage{color}
\usepackage{bm}

\definecolor{darkblue}{rgb}{0.0,0.0,0.55}
\RequirePackage[colorlinks,citecolor=darkblue,urlcolor=darkblue,linkcolor=darkblue]{hyperref} 
\usepackage[normalem]{ulem}

\usepackage[sort,comma]{natbib}

\setcitestyle{citesep={;}}

\allowdisplaybreaks

\newcommand{\rank}{{\rm rank}}
\newcommand{\Nor}{{\cal N}}  % for normal density
\newcommand{\ben}{\begin{enumerate}}
\newcommand{\een}{\end{enumerate}}
\newcommand{\beq}{\begin{equation}}
\newcommand{\eeq}{\end{equation}}

\newcommand{\argmin}{\operatornamewithlimits{argmin}}

\newcommand{\norm}[1]{\lVert#1\rVert}

\newcommand{\bx}{{\bf x}}

\newcommand{\by}{{\bf y}}

\newcommand{\defeq}{\operatorname{:=}}

\newcommand{\estHs}{\ensuremath{{\hat{\theta}}_{HS}}{}}

\DeclarePairedDelimiterX\MeijerM[3]{\lparen}{\rparen}%
{\begin{smallmatrix}#1 \\ #2\end{smallmatrix}\delimsize\vert\,#3}

\newcommand\MeijerG[8][]{%
  G^{\,#2,#3}_{#4,#5}\MeijerM[#1]{#6}{#7}{#8}}

\WithSuffix\newcommand\MeijerG*[7]{G^{\,#1,#2}_{#3,#4}\MeijerM*{#5}{#6}{#7}}
\DeclarePairedDelimiterX\pFqM[3]{\lparen}{\rparen}%
{\begin{smallmatrix}#1 \\ #2\end{smallmatrix}\delimsize\vert\,#3}

\newcommand\pFq[6][]{%
  {}_{#2}F_{#3}\pFqM[#1]{#4}{#5}{#6}}

\WithSuffix\newcommand\pFq*[5]{{}_{#1}F_{#2}\pFqM*{#3}{#4}{#5}}
\numberwithin{theorem}{section}

\numberwithin{Def}{section}

\numberwithin{remark}{section}

\numberwithin{proposition}{section}

\numberwithin{lemma}{section}

\numberwithin{Cor}{section}

\usepackage{booktabs,array}

\newcount\rowc

\makeatletter
\def\ttabular{%
\hbox\bgroup
\let\\\cr
\def\rulea{\ifnum\rowc=\@ne \hrule height 1.0pt \fi}
\def\ruleb{
\ifnum\rowc=1\hrule height 1.0pt  \else
\ifnum\rowc= 3  \hrule height 0.5pt \else%\heavyrulewidth 
\ifnum\rowc= 5  \hrule height 0.5pt \else%\heavyrulewidth 
\ifnum\rowc= 7  \hrule height 0.5pt \else%\heavyrulewidth 
\ifnum\rowc= 9  \hrule height 0.5pt \else%\heavyrulewidth 
\ifnum\rowc= 11  \hrule height 0.5pt %\heavyrulewidth 
  \else \hrule height 0pt%\lightrulewidth
\fi\fi\fi\fi\fi\fi}
\valign\bgroup
\global\rowc\@ne
\rulea
\hbox to 7em{\strut \hfill##\hfill}%
\ruleb
&&%
\global\advance\rowc\@ne
\hbox to 7em{\strut\hfill##\hfill}%
\ruleb
\cr}
\def\endttabular{%
\crcr\egroup\egroup}

\graphicspath{{../../figures/}{../figures/}{./figures/}{./}}

\defcitealias{wald1950statistical}{Wald's (1950)} 
\defcitealias{stein_inadmissibility_1956}{Stein's (1956)}

\title{\vspace{-1cm} \baselineskip=20pt  \bf Horseshoe Regularization for Machine Learning in Complex and Deep Models}
\date{}
\begin{document}
\maketitle
\baselineskip=15pt
\vspace{-1.75cm}
\begin{center}
 Anindya Bhadra \\
  Department of Statistics, Purdue University, 250 N. University St., West Lafayette, IN 47907, USA.\\
  bhadra@purdue.edu\\
  \hskip 5mm\\
   Jyotishka Datta \\
  Department of Mathematical Sciences, University of Arkansas, Fayetteville, AR 72704, USA.\\
  jd033@uark.edu\\
    \hskip 5mm\\
   Yunfan Li \\
    Department of Statistics, Purdue University, 250 N. University St., West Lafayette, IN 47907, USA.\\
    li896@purdue.edu\\
        \hskip 5mm\\
    Nicholas G. Polson\\
    Booth School of Business, University of Chicago, 5807 S. Woodlawn Ave., Chicago, IL 60637, USA.\\
    ngp@chicagobooth.edu\\
  \end{center}
\begin{abstract}
\noindent Since the advent of the horseshoe priors for regularization, global-local shrinkage methods have proved to be a fertile ground for the development of Bayesian methodology in machine learning, specifically for high-dimensional regression and classification problems. They have achieved remarkable success in computation, and enjoy strong theoretical support. Most of the existing literature has focused on the linear Gaussian case; see \citet{bhadra2017lasso} for a systematic survey. The purpose of the current article is to demonstrate that the horseshoe regularization is useful far more broadly, by reviewing both methodological and computational developments in complex models that are more relevant to machine learning applications. Specifically, we focus on methodological challenges in horseshoe regularization in nonlinear and non-Gaussian models;  multivariate models; and deep neural networks.  We also outline the recent computational developments in horseshoe shrinkage for complex models along with a list of available software implementations that allows one to venture out beyond the comfort zone of the canonical linear regression problems.
\end{abstract}

\section{Introduction}\label{sec:intro}

While Bayesian regularization is achieved through an appropriate choice of prior, many questions arise in designing sparsity priors in high-dimensional problems on both theoretical and computational fronts. Are the resulting posteriors ``optimal'' in some sense? Although the Bayesian posterior allows probabilistic uncertainty quantification, is it actually feasible to achieve a reasonable computational approximation of the posterior distribution in high dimensions? Fortunately, at least in the realm of linear Gaussian models, these questions are beginning to be answered in the affirmative over the past decade, thanks in large part due to the success of ``global-local'' priors, of which the horseshoe \citep{carvalho2010horseshoe} remains a canonical example. \citet{bhadra2017lasso} provide a detailed exposition of available results on global-local priors in linear models. They also provide extensive comparisons of global-local regularization with the lasso \citep{tibshirani96} in linear models, which is perhaps one of the most popular regularization methods. Yet, sparsity as a phenomenon is hardly limited to the prototypical but simplistic domain of the normal means model or the linear regression model with normal errors. Armed with the theoretical results and computational strategies developed for the horseshoe in linear models, what light could one shed, then, on the current state of the art in global-local shrinkage in nonlinear, non-Gaussian models?

The article is organized as follows. The remainder of Section~\ref{sec:intro} provides a brief historical overview of regularization, dating back to the work of \citet{stein_inadmissibility_1956}. Section~\ref{sec:gl} points out the critical differences between global and global-local regularization approaches in linear Gaussian models. Given this background, we proceed to the main focus area of our article: the current state of global-local regularization \emph{beyond} linear Gaussian models. While both Sections \ref{sec:shallow} and \ref{sec:deep} focus on these issues, we reserve the discussion of horseshoe shrinkage in \emph{shallow} models for the former and their more recent emerging uses in \emph{deep} models for the latter. Section~\ref{sec:app} describes the computational aspects of horseshoe shrinkage along with available software implementations in complex and deep models. Section \ref{sec:conc} concludes with some possible directions for future research.

\subsection{Regularization from a Bayesian perspective}\label{sec:reg}

Many penalized optimization problems in statistics are of the form 
\begin{eqnarray*}
\argmin_{\theta\in \mathbb{R}^n} \{l(\theta; y) + \lambda \pi(\theta)\},%\label{eq:pen}
\end{eqnarray*}
where $l(\theta; y) $ is a measure of fit of parameter $\theta$ to data $y$ (also known as the empirical risk), $\pi(\theta)$ is a penalty function and $\lambda$ is a tuning parameter. Let $p(y\mid \theta) \propto \exp\{-l(\theta; y)\}$ and $p(\theta) \propto \exp\{-\lambda \pi(\theta)\}$, where $p$ is a generic density. If $l(\theta;y)$ is proportional to the negative of the log likelihood function under a suitable model, one arrives at a Bayesian interpretation to the regularization problem: that of finding the mode of the posterior density $p(\theta \mid y )$ under prior density $p(\theta)$ \citep{polson2015mixtures}. The prior need not necessarily be proper but the posterior, $p(\theta \mid y) \propto p(y \mid \theta) p(\theta)$, may still be proper. This provides an equivalence between regularization and Bayesian methods. Common examples include the equivalence between ridge penalty and a Gaussian prior or the lasso penalty and a double exponential prior, when used in conjunction with a Gaussian likelihood which corresponds to a squared error loss.  We distinguish among the three following estimators throughout the paper.
\begin{enumerate}
\item $ \hat{\theta}_{mle} \defeq \argmin_{\theta\in \mathbb{R}^n}  l(\theta; y) $, the maximum likelihood estimator or mle, 
\item $\hat{\theta}_{mode} \defeq \argmin_{\theta\in \mathbb{R}^n} \{l(\theta; y) + \lambda \pi(\theta)\}$, the posterior mode,
\item $ \hat{\theta}_{mean} \defeq E(\theta \mid y) $, the posterior mean.
\end{enumerate}
While they are clearly connected, e.g., the mle is obtained as a special case of posterior mode if the prior is flat, early works in decision theory established only the third is guaranteed to be an admissible estimator under the squared error loss (at least, if the prior is proper), while the first two, in general, are not admissible.

\subsection{Decision theoretic foundations of Bayes regularization and global shrinkage}
\citetalias{wald1950statistical} search for an \emph{optimal} invariant procedure for multi-parameter models, similar in spirit to the UMVUE for single-parameter models, was dealt a major blow by the works of \citet{stein_inadmissibility_1956} and \citet{james_estimation_1961}. Formally, consider independent observations $y=(y_1,\ldots, y_n)$ from the model $(y_i\mid\theta_i) \sim \mathcal{N} (\theta_i, 1)$ where $\theta = (\theta_1,\ldots, \theta_n)$. Then, the James--Stein estimator 
\begin{align}
\hat{\theta}_{JS} &= \left(1 - \frac{n-2}{\sum_{i=1}^n y_i^2}\right) y \label{eq:JS},
\end{align}
has the property $R(\hat{\theta}_{JS} , \theta) < R( \hat{\theta}_{mle}, \theta)$ for all $\theta$ and all $n>2$, where $R(\delta, \theta)  =\mathbb{E}_{y\mid\theta} \norm{\delta - \theta}^2$. Thus, the mle of $\theta$, which in this case is just $y$ itself, is inadmissible.  The proposed \emph{biased} and \emph{nonlinear} estimator, which \emph{shrinks} the parameter vector to a pre-determined direction, itself turned out to be inadmissible, improved by the positive-part Stein estimator \citep{baranchik1970family} and several others since then. More than half a century and countless articles since Stein's pioneering result, \citet[][p.~102]{efron2016computer} assert \emph{``the main point here is that at present there is no optimality theory for shrinkage estimation. Fisher provided an elegant theory for optimal unbiased estimation. It remains to be seen whether biased estimation can be neatly codified.''} Although settling the question of optimality proved elusive, several researchers examined the far more modest, if not basic, requirement of admissibility, leading to a resurgence in Bayes procedures.  \citetalias{stein_inadmissibility_1956} risk result is frequentist, and makes no use of a prior. However, an empirical Bayes perspective was provided by \citet{robbins1956}, expanded further by \citet{efron1973stein}. Important connections with proper Bayes rules were established by \citet{strawderman1971proper},  who showed the existence of proper Bayes (hence, admissible) minimax estimators for all dimensions larger than or equal to five, and characterized the associated prior. Detailed decision theoretic perspectives can be found in \citet{berger1985statistical}, whereas a more recent book length treatment for shrinkage methods is by \citet{fourdrinier2017shrinkage}.

Although the James--Stein estimate strictly dominates the usual estimate $y$ in all dimensions larger than two, its performance can be vastly improved in high dimensions under an assumption of sparsity. More formally,  assume again the Gaussian sequence model $(y_i\mid\theta_i) \sim \mathcal{N} (\theta_i, 1)$, except $\theta$ now lies in the set $l_0[p_n] = \{\theta: \#(\theta_i \ne 0)\le p_n\}$ with $p_n = o(n)$ as $n\to\infty$, the case termed ``nearly-black'' by \citet{donoho1992maximum}. One example of a nearly-black parameter vector is the $r$-spike model considered by \citet{johnstone2004needles}, where $\theta$ contains $r$ non-zero components or spikes of magnitude $\sqrt{n/r}$, giving $\norm{\theta}^{2} = n$, with the remaining elements of $\theta$ set to zero. Under this model, the risk of the James--Stein estimator $\hat{\theta}_{JS}$ satisfies $R(\hat{\theta}_{JS}, \theta)  \ge n/2$, whereas if one is to simply threshold the observed data $y$ at $\sqrt{2\log n}$, the resulting estimator performs with risk $\sqrt{\log n}$ \citep{johnstone2004needles}. To understand this phenomenon more clearly, consider the form of the James--Stein estimator in Equation (\ref{eq:JS}). 
The term within the parentheses is the factor by which the observed data $y$ are shrunk. It is apparent that the shrinkage factor depends on the data only though the norm $\norm{y}^2$ and all terms are shrunk by the \emph{same factor}. Thus, the James--Stein estimator makes no distinction on whether the underlying $\theta_i$ is zero, in which case the corresponding  unbiased estimate $y_i$ should be aggressively shrunk towards zero to ensure an improvement in the overall mean squared error, or whether $\theta_i$ is non-zero and the corresponding $y_i$ should be shrunk as little as possible. We term this behavior \emph{global shrinkage}. Though this is entirely reasonable for the James--Stein estimate, which operates with no assumption of sparsity in $\theta$, recent interest in sparse models has led to a search for estimators that are more judicious on which terms to shrink.

\section{Horseshoe and the advent of global-local regularization}\label{sec:gl}
One of the earliest works to successfully distinguish between shrinking the noise terms while retaining the signal terms is the horseshoe estimator of \citet{carvalho2010horseshoe}. The estimator, $\hat\theta_{HS}$, is defined as the posterior mean of $\theta$ under the following hierarchical model (for $\sigma^2=1$):
\begin{align}
( y_i \mid \theta_i ) \sim \mathcal{N} ( \theta_i,1), \  ( \theta_i \mid \lambda_i, \tau ) \sim \mathcal{N} (0 , \lambda_i^2 \tau^2), \  \lambda_i  \sim C^+( 0 , 1),\ \tau \sim C^+( 0 , 1); \label{eq:hs}
\end{align}
where $X \sim C^+(0,1)$ denotes a standard half-Cauchy distributed random variable with density $p(x) = (2/\pi) (1 + x^2)^{-1}; \; x>0$. The first thing to notice is that in the hierarchical model all priors are proper and hence the resultant Bayes estimator is admissible. Second, the prior on $\theta_i$ is a scale mixture of normals with half-Cauchy mixing distributions. The marginal prior on $\theta_i$ is unbounded at the origin and has tails that decay polynomially \citep{carvalho2010horseshoe}. The ``global'' term $\tau$ is shared across all dimensions, while the $\lambda_i$ terms are component-specific, or ``local.'' The key intuition is that the term $\tau$ now adapts to the level of sparsity by typically settling on a small value, while the heavy-tailed $\lambda_i$ terms still allow the signals terms to escape from being shrunk too much. Along with this informal intuition, several theoretical and computational properties of global-local priors have been established, at least in the linear Gaussian setting. We summarize these properties separately.

\subsection{Theoretical properties in linear Gaussian models}\label{sec:theory}
Finite sample risk bounds for the horseshoe estimator were established by \citet{polson2012half} who showed (a) $\hat\theta_{HS}$ has a risk profile that is quite similar to the James--Stein estimate $\hat{\theta}_{JS} $ when $\norm{\theta}$ is large, and (b) $\hat\theta_{HS}$ offers a large benefit over  $\hat{\theta}_{JS}$ when $\norm{\theta}=0$. Both (a) and (b) were verified via simulations for various dimensions of $\theta$. When $\theta$ is sparse, a detailed theoretical understanding of the improvement over the James--Stein estimate in finite samples is given by \citet{bhadra2016prediction}, although in the context of prediction risk, rather than estimation risk. 

If one turns attention to asymptotic risk, rather than finite sample risk, several more results can be found. Some of the more prominent ones are as follows.
\begin{enumerate}
\item \cite{carvalho2010horseshoe} studied information-theoretic properties of the horseshoe estimator when the true parameter vector is sparse. They obtained a better upper bound on the asymptotic Kullback--Leibler risk of the posterior predictive density with respect to the true density of the data generating model compared to any other prior density that is bounded above at the origin.
\item \cite{datta2013asymptotic} proved that the decision rule induced by the horseshoe estimator is asymptotically Bayes optimal for multiple testing under 0--1 loss up to a multiplicative constant. This result was generalized to include other global-local priors by \citet{ghosh2016asymptotic} and \citet{bai2018large}, among others.
\item \cite{van2014horseshoe} showed the horseshoe estimator is minimax in $\ell_2$ in a nearly-black case up to a constant. Specifically,
$$ 
\sup_{ \theta \in l_0[p_n] } \; 
\mathbb{E}_{ y | \theta } \norm{ \estHs (y) - \theta }^2 \asymp
p_n \log \left ( n / p_n \right ),
$$
which is the asymptotic minimax rate when $\theta \in l_0[p_n]$, established by \citet{donoho1992maximum}. Here $a_n \asymp b_n$ denotes $\lim_{n\to\infty} a_n/b_n=c\in (0,\infty)$. The result was expanded by \citet{van2015conditions} and \citet{ghosh2017asymptotic} to prove several other priors, such as the horseshoe+ \citep{bhadra2015horseshoe+}, the normal-gamma \citep{griffin2005alternative} and the spike-and-slab lasso \citep{rovckova2016spike}, also result in asymptotic minimax estimates.
\item Turning attention to uncertainty quantification, \citet{vanderpas2017} proved that the posterior credible intervals under the horseshoe prior also have good frequentist coverage properties in an asymptotic sense, provided the choice of the global shrinkage parameter $\tau$ meets certain restrictions.
\end{enumerate}

\subsection{Computational properties in linear Gaussian models}
Before the widespread popularity of global-local shrinkage approaches, traditional sparse Bayesian models relied on the so-called \emph{spike-and-slab} priors \citep{mitchell88} where conditionally i.i.d. $\theta_i$ are modeled as
\beq
\theta_i \mid p, \psi  \sim (1-p)\delta_{\{0\}} + p \Nor (0, \psi^2), \label{spikeslab}
\eeq
a two-component mixture, where $\delta_{\{0\}}$ denotes a point mass at zero. The model is in many senses natural; it reflects the prior belief of a Bayesian that a parameter $\theta_i$ has non-negligible probability $(1-p)$ of being zero. The model also has many attractive theoretical properties, including asymptotic optimality in testing under 0-1 loss \citep{bogdan2011asymptotic}, and asymptotic minimaxity in estimation under $\ell_2$ loss when a Laplace or heavier tailed prior is used for the slab distribution \citep{castillo2012needles}. Although the spike-and-slab model produces exact zero estimates, the chief difficulty under this prior is in exploring the posterior. While posterior means or quantiles can be found in polynomial time algorithms under a spike-and-slab model \citep{castillo2012needles}, exploring the entire posterior incurs extreme computational cost, primarily since there is no good way to avoid sampling the binary indicators denoting whether a parameter is zero versus non-zero, and this in turn leads to a combinatorial problem. While significant advances have been made in finding posterior \emph{point estimates} such as posterior modes using an expectation-maximization (EM) algorithm under a relaxed continuous spike-and-slab model \citep{rovckova2014emvs} or its variants such as the spike-and-slab lasso model \citep{rovckova2016spike}, comparative studies of full posterior exploration by Markov chain Monte Carlo (MCMC) techniques have indicated that global-local priors offer significant computational benefits over point mass spike-and-slab mixture priors \citep{li2017variable}. Moreover, posterior modes under global-local priors are also available using fast EM type algorithms and have been shown to be computationally and statistically quite competitive to frequentist counterparts such as the lasso \citep{bhadra2017horseshoe}. We elaborate further on computational aspects in Section~\ref{sec:app}, where we also list available software implementations of global-local shrinkage approaches known to us.

\section{Horseshoe shrinkage in shallow nonlinear, non-Gaussian models}\label{sec:shallow}

We define models as shallow where the parameter of interest lies one level below the observed data model, although the prior model for the parameter of interest might itself contain a number of levels of hierarchy. Thus, the normal means model $(y_i \mid \theta_i) \sim \Nor(\theta_i, 1)$ and the linear regression model $(y \mid X, \theta) \sim \Nor (X\theta, 1)$ fall in this framework, regardless of the number of hierarchies used in defining the prior on $\theta$, whereas a deep neural network model does not necessarily fall in this framework. Nevertheless, several advances have been made in shallow models other than the normal means or linear regression models of Section~\ref{sec:gl}, and we point out a few important developments below.

\subsection{Shallow models with Gaussian errors}

\textbf{Nonlinear function estimation and default Bayes analysis:} An early work on the use of horseshoe and horseshoe+ priors in estimating low-dimensional functions of the high-dimensional normal means model of Section~\ref{sec:intro} is by \citet{bhadra2015default}. They consider the following four one dimensional functions of $\theta$: $\psi_1 = \sum_{i=1}^{n} \theta_i^2, \psi_2 = \max_{1\le i\le n} \theta_i, \psi_3 = \theta_1\theta_2$ and $\psi_4 = \theta_1/\theta_2$, where for $\psi_3$ and $\psi_4$ the remaining $\theta_i$s are nuisance parameters. They demonstrate that using the horseshoe and horseshoe+ priors on $\theta$ enables non-informative Bayesian analysis in each of these problems, resolving a long-standing paradox by \citet{efron1973discussion}, who pointed out the difficulty in designing a prior on $\theta$ that simultaneously enables non-informative analysis in all four problems. The key contribution of \citet{bhadra2015default} is to demonstrate the regularly varying tails of global-local horseshoe priors on $\theta$ translate to the induced priors in each of the four nonlinearly transformed parameters above, which preserve regular variation of the prior. \citet{bhadra2015default} then appeal to the relative tail heaviness of the prior and the likelihood considered by \citet{dawid1973posterior} to explain the non informative Bayes answer. The performance of the global-local priors is quite competitive to the reference priors \citep{bernardo1979reference, berger_development_1992} for these problems, when the reference priors exist. An added benefit is that the horseshoe priors are proper, thereby circumventing model selection problems encountered with improper reference priors.\\

\noindent \textbf{Nonparametric function estimation:} \citet{shin2016functional} consider a standard nonparametric regression model with additive Gaussian noise of the form $Y_i= F(x_i) + \epsilon_i$, for $i=1,\ldots,n$, where i.i.d. $\epsilon_i \sim \Nor(0,\sigma^2)$ and $F(x) = E(Y\mid x)$. A natural representation of $F$ is by a basis expansion of the form $f(x) =  \phi(x)^{T}\beta$, appealing to  Karhunen--Lo\`eve representation, where $\phi = (\phi_1, \ldots, \phi_k)$ is a suitable choice of basis functions. The problem is then to estimate $\beta= (\beta_1,\ldots, \beta_k)^{T}$. The interpretation of sparsity in $\beta$ is somewhat more subtle, however. While it is certainly possible to put a sparsity prior such as horseshoe directly on $\beta$, it is not clear what a global-local shrinkage of certain basis coefficients exactly achieves. If one possesses a prior belief on the shape of $F$ as being close to a certain parametric family (e.g., linear or quadratic), perhaps it is more reasonable to shrink toward that shape. Define $\phi_0$ to be the column space of the parametric function one desires to shrink to. For example, if the parametric form is assumed to be close to linear then $\phi_0 = \{1,x\}\in \mathbb{R}^{n\times 2}$. To this end, \citet{shin2016functional} propose the ``functional horseshoe'' prior on $\beta$, with the prior density given as
\begin{align*}
p(\beta \mid \tau) &\propto (\tau^2)^{-(k - d_0)/2} \exp\left\{-\frac{1}{2\sigma^2 \tau^2} \beta^{T}\phi^{T} (I - Q_0)\phi \beta\right\},\\
p(\tau)&\propto \frac{(\tau^2)^{b-1/2}}{(1+\tau^2)^{(a+b)}};\; \tau, a, b >0,
\end{align*}
where $d_0 = \rank(\phi_0)$ and $Q_0 = \phi_0 (\phi_0^T\phi_0)^{-1}\phi_0^T$, is the projection matrix of $\phi_0$. The marginal prior on $\beta$ can be seen to a normal scale mixture and the prior on $\tau$ is half-Cauchy when $a=b=1/2$. The term $(I-Q_0)$ in the prior inverse covariance enables shrinkage of $\phi$ towards $\phi_0$. In effect, this formulation imposes a non-informative prior on the coefficients of the base parametric model. \citet{shin2016functional} establish consistency of model selection under this prior and demonstrate good empirical results. \\

\noindent \textbf{Dependent data models:} Horseshoe priors have also been considered for dynamic process models for time series data. \citet{kowal2017dynamic} define a log-scale representation of the variance term  in Equation (\ref{eq:hs}) as $h_i= \log (\tau^2 \lambda_i^2)$ and point out that the marginal horseshoe hierarchy up to $(\theta_i \mid \tau)$ is obtained by the following model
$$
h_i = \mu + \eta_i,\; \eta_i \sim Z(1/2, 1/2,0,1),
$$
where $\mu = \log(\tau^2), \eta_i = \log (\lambda_i^2)$ and $Z$ denotes the Fisher-Z distribution \citep{barndorff1982normal}. Next, they introduce dependence in $h_i$s using an autoregressive structure as
$$
h_{i} = \mu + \phi(h_{i-1} - \mu) + \eta_i,\; \eta_i \sim Z(1/2, 1/2,0,1),
$$
where the model has one additional parameter $\phi$, which controls the strength of dependence. For sampling,  \citet{kowal2017dynamic} rely on the normal mean-variance mixture representation of Fisher-Z random variables with respect to P\'olya-gamma mixing density \citep{polson2013bayesian} and derive a computationally efficient Gibbs sampler., with an application in Bayesian trend filtering. A closely related formulation is used by \citet{bitto2018achieving}, who use a double gamma prior on the variance terms in a dynamic linear model for achieving shrinkage, instead of putting a prior on the log of the variance terms. \\

\noindent \textbf{Graphical models:} Moving away from univariate Gaussian error models, \citet{li2017graphical} consider the multivariate Gaussian model $y_i \mid \Omega \sim \Nor(0,\Omega^{-1})$ for $i=1,\ldots, n$, where $\Omega$ is a $p\times p$ inverse covariance matrix such that $p>n$. In this setting, off-diagonal zeros in $\Omega$ encode conditional independence among variables \citep{lauritzen1996graphical}. The model is a fundamental building block in network analysis, and has numerous applications in genomics, econometrics and virtually every other field where network data are encountered. To achieve sparsity in $\Omega=\{\omega_{ij}\}; i, j=1,\ldots, p$; \citet{li2017graphical} assume the following prior, which they term the ``graphical horseshoe:''
$$
\omega_{ii} \propto 1,\; \omega_{ij,i<j} \sim \Nor(0, \lambda_{ij}^2 \tau^2),\; \lambda_{ij, i<j} \sim C^{+}(0,1),\; \tau \sim C^{+} (0,1),
$$
with the prior mass truncated to the space of positive definite matrices. That is, they assume a non-informative prior on the diagonal $\omega_{ii}$ terms and independent horseshoe priors on off-diagonal $\omega_{ij}$ terms to induce a sparse graphical structure. The sampling schemes outlined in \citet{wang2012bayesiangl} and \citet{makalic2016samplerHS} are used to design a computationally efficient Gibbs sampler that maintains positive definiteness of the posterior estimate of $\Omega$. The resulting estimate compares favorably with respect to popular alternatives, such as the graphical lasso \citep{friedman2008glasso} or the graphical SCAD \citep{lam2009sparsistency}. It is worth noting similar models have been developed using the spike-and-slab lasso type priors, which rely on fast EM approaches for finding the posterior mode \citep[e.g.,][]{deshpande2017simultaneous,li2018bayesian} , but it remains a challenge to perform efficient posterior exploration under these priors. 

An alternative approach for sparse inverse covariance estimation is developed by \citet{williams2018bayesian}, who use as a starting point the partial regression equations in a multivariate Gaussian model, given for $i=1,\ldots,p$ by
$$
y_i =  y_{-i}\beta_{-i} + \epsilon_i,
$$
where $\epsilon_i \sim \Nor(0, \omega_{ii}^{-1})$ and $\omega_{ij}/\omega_{ii} = -\beta_{ij}$. Independent horseshoe priors are then imposed on the $\beta_{ij}$ terms to enable a sparse estimation of $\Omega$. A potential drawback of this approach is that under independent horseshoe priors on the partial regression coefficients, one does not necessarily obtain a symmetric and positive definite estimate of $\Omega$. Thus, one needs to follow a neighborhood selection approach \citep{meinshausen2006high} to perform covariance selection, followed by a symmetrization step.\\

\noindent \textbf{Seemingly unrelated regression models:} Seemingly unrelated regression models concern regressing multiple correlated responses on multiple predictors, where the error terms display a covariance structure. It is often of interest to simultaneously infer the regression coefficients and the error precision matrix. The model is quite flexible, with the multiple linear regression model and zero mean graphical models as special cases. \citet{li2019joint} consider the linear model $Y_{n\times p} = X_{n\times q} B_{q\times p} + E_{n\times p}$, where the error term is assumed to follow a matrix normal distribution \citep{dawid1981some}, i.e., $E\sim \mathrm{MN}_{n\times p} (0, I_n, \Omega_{p\times p}^{-1})$. The problem is then a joint estimation of $B$ and $\Omega$. Early works by \citet{zellner1962efficient} demonstrated one incurs a loss of efficiency if the error covariance structure is not accounted for while estimating $B$. However, these early results are ill-suited to handle modern applications such as genomic data analysis where both the number of features $q$ and number of responses $p$ routinely exceed the sample size $n$. To enable a sparse estimation, \citet{li2019joint} assume independent horseshoe priors on $B_{ij}$ and the graphical horseshoe prior of  \citet{li2017graphical} on $\Omega$. A fully Bayesian estimation algorithm is proposed that is linear in $q$, and improved statistical performance over competing approaches using spike-and-slab type priors \citep[e.g.,][]{bhadra2013joint} is demonstrated in simulations.\\

\noindent \textbf{Factor models:} A factor model offers a particular low rank decomposition for modeling a $p\times p$ covariance matrix as $\Sigma = BB^{T} + \Psi^2$ where $B\in \mathbb{R}^{p\times k}$ with $k<p$ is the matrix of factor loadings and $\Psi^2$ is a diagonal matrix with non-negative elements. Any covariance matrix allows this decomposition and when $k < p$, the model offers an effective method for low-rank decomposition to facilitate estimation when the sample size is small. Identifiability conditions under this model are well-studied. \citet{hahn18} applied horseshoe priors on the factor loadings and developed an efficient elliptical slice sampler for posterior sampling. The shrinkage achieved by the horseshoe priors allows one to incorporate many instruments into the analysis.

\subsection{Shallow models with non-Gaussian errors}

\noindent \textbf{Gamma--Poisson glms:} The use of global-local priors has also been extended to generalized linear models (glm), a common tool for modeling data distributed according to exponential family distributions, not necessarily Gaussian. An early example is by \citet{datta16}, who use global-local priors in a quasi-sparse gamma-Poisson glm. Their model is:
$$
y_i \mid \theta_i \sim \mathrm{Poisson} (\theta_i),\quad \theta_i \mid \lambda_i, \tau \sim \mathrm{Gamma}(\alpha, \lambda_i^2\tau^2),
$$
with heavy-tailed prior densities on $\lambda_i$ and $\tau$, where half-Cauchy is one choice. This parameterization encourages both an abundance of zero and small non-zero counts, as well as large counts; an analog to sharp spike at zero and heavy tails for real-valued data in the usual implementation of the horseshoe prior. The method proved successful in detecting rare mutations in a massive genetic sequencing data set, where the observations are counts; and signals, indicating presence of mutations, are rare.\\

\noindent \textbf{Classification using probit and logistic models:} An important application of multi-class classification in machine learning is in topic modeling, where given a text, one tries to map it into belonging to one of several pre-defined set of topics or classes. \citet{magnusson2016dolda} combine the diagonal orthant (DO) probit model of \citet{johndrow2013diagonal} with latent Dirichlet allocation (LDA) \citep{blei2003latent, Pritchard945}, resulting in a method they call DOLDA. Here the horseshoe prior is used to achieve shrinkage of the coefficients in the probit model and results in substantially better accuracy of topic prediction over standard approaches in topic modeling. Some theoretical support for using shrinkage priors on regression coefficients is provided by \citet{wei2017contraction}, who derive posterior contraction rate similar to that of point mass mixture priors \citep{atchade2017contraction}, which are shown to be better than the Bayesian lasso or non-informative normal priors on the regression coefficients.  Notable computational success is achieved by \citet{terenin2018gpu}, who point out the latent variables in a probit model are conditionally independent of each other, and consequently, can be sampled in parallel in a massively multithreaded environment, such as using a graphics processing unit (GPU). They use a case study of the horseshoe probit regression where a million data points in several thousand dimensions are classified in several minutes of running time. \\

\noindent \textbf{Arbitrary glms via Gaussian approximation:} Another approach is developed by \citet{piironen2016hyperprior} for arbitrary glms, where inference proceeds via dynamic Hamiltonian Monte Carlo and is implemented in the probabilistic programming language Stan \citep{carpenter2017stan}. To set the global scale parameter $\tau$, the method estimates the number of non-zero components in the model by approximating the glm likelihood function via a Gaussian likelihood, where the first two moments of the Gaussian approximation are calculated via score matching using a second order Taylor approximation of the actual glm likelihood. The success of this method, of course, depends in part on the appropriateness of the Gaussian approximation of the likelihood function while estimating the number of zero components. \citet{piironen2016hyperprior} demonstrate good empirical performance in classification problems under a logistic model using four different real data sets. 

\section{Horseshoe shrinkage in deep models}\label{sec:deep}

The use of horseshoe and other regularizing priors have also made their way into \emph{deep} models, where the parameters of interest share multiple levels of hierarchy below the level of the observed data. Yet, to our knowledge, the literature here is far more sparse compared to the use of global-local shrinkage in \emph{shallow} models, described in Section~\ref{sec:shallow}. Indeed, this is surprising, since many prototypical deep models, such as deep neural networks, are heavily overparameterized with respect to what is actually observed and therefore, deep models should be a fertile ground for the application sparsifying priors. The chief difficulty appears to be computational.  Nonetheless, the goal of the current section is to summarize the existing works.\\

\noindent \textbf{Horseshoe shrinkage in deep neural networks:} \citet{ghosh17} and \citet{ghosh2018structured} apply horseshoe priors for model selection in deep Bayesian neural networks. A related work is by \citet{louizos2017bayesian} who use group horseshoe priors for regularizing the weights in a deep neural network. A unifying feature of these works is that they deploy Bayesian alternatives to the successful ``dropout'' mechanism \citep{srivastava2014dropout} for regularizing highly overparameterized deep neural networks. Dropout involves introducing multiplicative Bernoulli distributed noise variables into the hidden layers of a deep neural network in order to \emph{zero out} certain weight parameters, thereby encouraging sparsity. The connection between dropout and structured shrinkage regularization (including horseshoe) in deep models is established by \citet{nalisnick2018unifying}, who show the two forms of regularization can be obtained from each other via reparameterization. An important difference with shallow models is that full Bayesian computational inference via MCMC is considerably more  challenging in deep models. Consequently, inference usually proceeds using a variational approximation to the true posterior, with the approximating distribution usually being normal or belonging to some other family for which the parameters can be optimized efficiently. We provide some technical details below.

Given a set of weights $W_1,\ldots, W_L$, a set of biases $b_1,\ldots,b_L$ and input $\bx$, the output of a deep neural network with $L-1$ hidden layers is characterized by
$$
\by = (f_{W_1, b_1}\odot\ldots \odot f_{W_L,b_L})(\bx),
$$
where $\odot$ denotes the composition operation and $f_{W_l,b_l} (\bx)= f(b_l + W_l \bx)$, for a nonlinear activation function $f(\cdot)$. Thus, the network repeatedly applies a linear transformation to the inputs at each layer, before passing it through a nonlinear activation function. For the purpose of this article, we omit the bias terms henceforth. Here $W_l$ is a matrix of size $K_l\times K_{l+1}$, where $K_l$ is the number of units in layer $l$, with the final layer denoting the output layer. In this formulation,  \citet{ghosh17} and \citet{ghosh2018structured} assume horseshoe priors on the networks weights of the form
\begin{align}
( w_{kl} \mid  \lambda_{kl}, \tau_l ) \sim \mathcal{N} (0 , \lambda_{kl}^2 \tau_l^2 \mathbb{I}), \  \lambda_{kl}  \sim C^+( 0 , 1),\ \tau_l \sim C^+( 0 , 1); \label{eq:nn}
\end{align}
where $w_{kl}$ is a vector of all weights connected to unit $k$ in layer $l$ and $\mathbb{I}$ denotes the identity matrix. This formulation allows different global shrinkage parameters $\tau_l$ for different layers, while still admitting the usual global-local shrinkage of network weights within the same layer. 

\citet{nalisnick2018unifying} point out that the scale mixture representation in Equation (\ref{eq:nn}) can also be written as 
\begin{align}
(w_{kl} \mid  \lambda_{kl}, \tau_l ) \sim \lambda_{kl} \tau_l \mathcal{N} (0 , \mathbb{I}),\label{eq:non}
\end{align}
with the same prior distributions on $\lambda_{kl}$ and $\tau_l$. This closely parallels the dropout formulation of \citep{srivastava2014dropout}, where the multiplicative terms to the weights are binary indicators, causing some $w_{kl}$ terms to be exactly zero. In this sense, the horseshoe regularization can be understood to provide a continuous relaxation to exact sparsity in deep models, closely paralleling the situation with global-local versus spike-and-slab priors in linear models.

When the output of the neural network is binary, further data augmentation is needed to leverage the global-local priors. A notable work is by \citet{gan2015learning} who used the three parameter beta shrinkage \citep{armagan2011generalized} in a deep sigmoidal belief network. The P\'olya--gamma data augmentation scheme was used for augmenting the logistic likelihood in a sigmoidal belief network resulting in conjugate models for efficient sampling, using the strategy identified by \citet{polson2013bayesian} for shallow logistic regression models. \citet{wang2019scalable} expand upon this line of work, identifying data augmentation strategies for deep neural networks under different choices of nonlinear activation functions, such as ReLU, logistic and hinge loss. A notable feature of \citet{wang2019scalable} is that they specify a completely simulation-based strategy for optimizing the parameters in a deep neural network (at least, for the output layer), following the MCMC-MLE technique of \citet{jacquier2007mcmc}, thereby bypassing the need for gradient-based training methods. 

 \citet{nalisnickautomatic} use the horseshoe prior to regularize both the number of hidden units per layer and the number of layers in a deep neural network, giving rise to the Automatic Relevance Detection--Automatic Depth Determination (ARD--ADD) prior. Their prior hierarchy is the same as in Equation (\ref{eq:nn}), except they use the improper prior $p(\tau_l)\propto 1/\tau_l$, which offers a stronger pull towards zero compared to a half-Cauchy prior on $\tau_l$. Inference proceeds via a variational EM algorithm and performance of the horseshoe prior is shown to be quite competitive to other traditional regularization mechanisms, such as dropout.

Progress has also been made in designing sparsity-inducing priors inspired by neural network architectures. A recent example is by \citet{shin2018neuronized}, who introduce ``neuronized priors.'' Their priors on $\theta_j$ for the Gaussian sequence model are of the form 
$$
\theta_j = T(\alpha_j - \alpha_0)w_j,
$$
where $\alpha_0$ is a constant, $\alpha_j \sim N(0,1)$, $w_j \sim N(0, \tau_w^2)$ and $T$ is a non-decreasing activation function. \citet{shin2018neuronized} show that certain choices of activation functions correspond to special cases of sparsity priors. For example, with the ReLU activation, i.e., $T(x)=\max(x,0)$, the induced prior on $\theta_j$ is the discrete spike-and-slab prior. Similarly, the horseshoe prior arises through another suitable choice of $T(\cdot)$. An advantage of this formulation is that a unified MCMC sampling strategy and fast EM algorithms for exploring the posterior mode can be built for a broad class of priors. \\

\noindent \textbf{Horseshoe shrinkage in deep glms:} The output of a traditional deep neural network is either real valued, when used in regression problems; or categorical, when the neural network is trained to perform classification. Deep models have also been proposed for glms, which is a flexible technique for modeling data distributed according to exponential family distributions, and can model a large class of response variables, including real-valued, categorical or counts. \citet{tran2018bayesian} develop flexible versions for glms using the output of a feedforward neural network. With responses $y$ and predictors $X$, a conventional glm models $E(y\mid X) = g(X\beta)$, where $\beta$ are the regression coefficients and $g(\cdot)$ is the link function. Thus, the conditional mean of the responses is a linear function of $X$, transformed through the link function $g$. This  linearity assumption is often restrictive and a natural way to introduce nonlinearity  is by replacing $X$ with the output of a multi-layer feedforward neural network that has $X$ as input and consequently, whose output is a nonlinear function of $X$.  \citet{tran2018bayesian} term this model DeepGLM. Similar to deep neural networks, a variational approximation to the log likelihood is used for training the model and global-local priors are used for inducing sparsity on $\beta$.

\section{Computational aspects and software implementations}\label{sec:app}
The existing implementations of horseshoe shrinkage for the models described in the previous two sections can be broadly categorized as fully Bayesian and approximately Bayesian. While exploring the entire posterior might be desirable viewed through a lens of Bayesian orthodoxy, the computational burden is often great enough to compel a researcher to explore suitable alternatives. At present, this effect is particularly acute for deep models, and almost all approaches attempting a Bayesian analysis for such models that are known to us rely on variational approximation or point estimation techniques. Nevertheless, we summarize fully and approximately Bayesian approaches for some important models in this section, along with a list of available software implementations. The associated issue of hyper-parameter selection is discussed at length in Section 5 of  \citet{bhadra2017lasso}.

\subsection{Fully Bayesian approaches in linear models}
Fully Bayesian approaches are more computation-intensive compared to point estimation approaches almost by design, since the focus is typically on approximating the full posterior. As a starting point, consider the linear regression model $y=X\beta + \epsilon$ where $y\in\mathbb{R}^n$ and $X\in \mathbb{R}^{n\times p}$. Assume one decides to use the horseshoe prior on $\beta$, i.e., $\beta_i \sim \Nor(0,\lambda_i^2\tau^2)$. The full conditional posterior of $\beta$ under this model is $\beta \mid \mathrm{rest} \sim \Nor((X'X+ \Lambda^{-2}\tau^{-2})^{-1}X' y, (X'X+ \Lambda^{-2}\tau^{-2})^{-1})$, where $\Lambda = \mathrm{diag}(\lambda_i)$. The computational bottleneck is in inverting the $p\times p$ matrix $(X'X+ \Lambda^{-2}\tau^{-2})$, with complexity $O(p^3)$, which is prohibitive when $p$ is large. Thus, posterior exploration even under this very simple model is challenging. A breakthrough is by \citet{bhattacharya_fast_2015}, who designed an exact algorithm for sampling from this model with complexity $O(n^2p)$, an improvement by two orders of magnitude when $n \sim o(p^{0.5})$, a situation which is commonly encountered in many high-dimensional applications. The key to their innovation is to replace the matrix inversion step by the solution to a system of linear equations. A more recent work by \citet{johndrow2017scalable} claims to reduce the complexity further to $O(\min\{s^3, sn\})$, where $s = \#\{i: \tau^{2} \lambda_i^{2} > \delta\}$ for some pre-defined threshold $\delta$. The algorithm is not exact but is demonstrated to have good empirical performance and enjoys some theoretical support. The advantage here is that $s$ is typically much smaller than $p$. \citet{nishimura2018prior} discuss the use of conjugate gradient methods with appropriate pre-conditioning for the same problem, with the focus again being on avoiding a costly matrix inversion.

While the strategies above focus on the posterior sampling of $\beta$, the global and local shrinkage parameters $\lambda_i$ and $\tau$ also need to be sampled. The early works in global-local shrinkage \citep[e.g.,][]{carvalho2010horseshoe, scott2010parameter} relied on slice sampling approaches. A more recent elliptical slice sampling approach for horseshoe priors, but on a transformed coordinate system (i.e., polar rather than Cartesian) is by \citet{hahn2018efficient}. However, \citet{makalic2016samplerHS} established that fully conjugate sampling of all parameters is possible in a linear regression model, with horseshoe priors on $\beta$. The key to their innovation is the observation that if $X^2 \mid a \sim \mathrm{IG}(1/2, 1/a)$ and $a\sim \mathrm{IG} (1/2, 1/k^2)$ then marginally, $X\sim C^{+}(0,k)$. Here $\mathrm{IG}(\alpha,\beta)$ denotes an inverse gamma random variable with shape parameter $\alpha$ and scale parameter $\beta$. The inverse gamma distribution is conjugate to itself, and to the variance parameter in a linear regression model with normal errors. Thus, this data augmentation strategy applied to $\lambda_i$ and $\tau$ allows for conjugate updates, thereby aiding the design of a Gibbs sampler, with the benefits of automatic tuning and no sample rejection. This augmentation strategy also plays a key role in the development of variational Bayes approaches, discussed in Section~\ref{sec:varbayes}.

%The main idea is to write the half-Cauchy density as a mixture of two inverse gamma random variables, which then allows conjugate updates for $\lambda_i$ and $\tau$. This formulation allows Gibbs samplers for global-local approaches, with the benefits of automatic tuning and no sample rejection.

\subsection{Fully Bayesian approaches in multivariate and Non-Gaussian models}\label{sec:fullbayes}
\citet{li2017graphical} consider i.i.d. observations from the $p$-variate normal model $y_i \mid \Omega \sim \Nor(0,\Omega^{-1}),$ for $i=1,\ldots,n$ and assume the following hierarchy, which they term the graphical horseshoe (GHS):
$$
\omega_{ii} \propto 1,\; \omega_{ij,i<j} \sim \Nor(0, \lambda_{ij}^2 \tau^2),\; \lambda_{ij, i<j} \sim C^{+}(0,1),\; \tau \sim C^{+} (0,1).
$$
Estimation of the inverse covariance matrix presents the additional complication that the estimate needs to be symmetric and positive definite. This is accomplished by combining the variable transformation technique first identified in the context of the Bayesian graphical lasso by \citet{wang2012bayesiangl} with the data augmentation scheme of  \citet{makalic2016samplerHS}. Let $S=Y'Y$ and partition the matrices $\Omega, S$ and $\Lambda$ as:
	\begin{align*}
	   \Omega=
	   \left( {\begin{array}{cc}
	   	\Omega_{(-p)(-p)} & \bm{\omega}_{(-p)p} \\  \bm{\omega}'_{(-p)p} & {\omega}_{pp} \  \end{array} } \right), \
	   S=
	   \left( {\begin{array}{cc}
	    S_{(-p)(-p)} & \mathbf{s}_{(-p)p} \\  \mathbf{s}'_{(-p)p} & {s}_{pp} \      \end{array} } \right), \
	   \Lambda=
	   \left( {\begin{array}{cc}
	   	\Lambda_{(-p)(-p)} & \bm{\lambda}_{(-p)p} \\  \bm{\lambda}'_{(-p)p} & 1 \  \end{array} } \right), \	
	\end{align*}
	where $(-p)$ denotes the set of all indices except for $p$, and $\Lambda_{(-p)(-p)}$ and $\bm{\lambda}_{(-p)p}$ have entries $\lambda_{ij}^2$. Diagonal elements of $\Lambda_{(-p)(-p)}$ can be arbitrarily set to 1. The key contribution of \citet{wang2012bayesiangl} is to show under the reparameterization $\bm{\beta}=\bm\omega_{(-p)p}$ and $\gamma={\omega}_{pp}-\bm{\omega}_{(-p)p}'\Omega_{(-p)(-p)}^{-1}\bm{\omega}_{(-p)p}$, the full conditional posteriors of $\bm\beta$ and $\gamma$ are available in closed form as a multivariate normal and a univariate gamma random variable, respectively. Moreover, this reparameterization also maintains positive definiteness of the posterior estimate. Combining this observation with the \citet{makalic2016samplerHS} scheme establishes all required conditional in the GHS model either as normal, gamma or inverse gamma, resulting in a Gibbs sampler. The computational complexity is $O(p^3)$.
	
\citet{li2019joint} take this approach a step further to consider the seemingly unrelated regression model $(y_i \mid X_i, \beta, \Omega) \sim \Nor(X_i\beta,\Omega^{-1})$ where $y_i\in\mathbb{R}^p$ and $X_i \in \mathbb{R}^q$ for $i=1,\ldots,n$ and both $\beta\in \mathbb{R}^{q\times p}$ and $\Omega \in \mathbb{R}^{p\times p}$ are unknown. Horseshoe and graphical horseshoe priors are used on $\beta$ and $\Omega$ respectively and once again, combing the methods of \citet{bhattacharya_fast_2015},  \citet{makalic2016samplerHS} and \citet{li2017graphical}, all updates are in closed form and the method only requires sampling from a multivariate normal or univariate gamma distributions. The computational complexity is $O(n^2 q p^3)$, making this the first fully Bayesian estimation algorithm in a joint mean--covariance estimation problem with a complexity linear in $q$, the number of covariates.

Computational approaches for some other nonlinear problems, such as the functional horseshoe prior of \citet{shin2016functional} proceed similarly to the linear regression case, with the covariates $X$ replaced by the basis functions. Similarly, the estimation of latent factors by \citet{hahn18} also relies on the slice sampling technique developed for linear models. We do not discuss them in detail. Computational approaches for glms are also similar to the linear models, except one now has to account for the link function. A standard technique is to use a data augmentation scheme for conjugate sampling in the posterior, which may be used in conjunction with global-local priors on regression coefficients. Examples include augmentation by a latent P{\'o}lya-gamma random variable for logistic regression models \citep{polson2013bayesian}, or by a latent Gaussian random variable for probit models \citep{albert1993bayesian}, with a unified framework for data augmentation for global-local priors described by \citet{bhadra2016global}. A similar technique for non-Gaussian regression using normal mean-variance mixtures is described by \citet{polson2013data} and the proposed approach has recently found use in modeling the nonlinearly transformed output layer in a multi-layer feedforward neural network \citep{wang2019scalable}. When used in conjunction with a sparsifying prior such as the horseshoe, these data augmentation techniques act as fully Bayesian analogs to the successful dropout mechanism for regularizing deep neural networks. Nevertheless, even with efficient Gibbs sampling schemes, fully Bayesian MCMC approaches are still computationally prohibitive, especially in deep models, where the number of parameters increases exponentially as a function of depth. This has led researchers to seek out computationally scalable alternatives, which we describe next.

\subsection{Variational Bayes and point estimation approaches}\label{sec:varbayes}
Variational Bayes methods work by replacing the true posterior distribution by an approximating distribution where the approximating distribution typically belongs to a simple parametric family that is easy to optimize. The best approximating distribution, within its class, is chosen by minimizing some measure of divergence (usually, the Kullback--Leibler divergence) with respect to the true density one is trying to approximate. This offers the added benefit that the normalizing constant for the approximating density is usually available in closed form, whereas it might be intractable for the true posterior. The downside of course is that variational Bayes methods can result in inconsistent estimates, even in relatively simpler settings such as state space models \citep{wang2004lack}. Nevertheless, the considerable computational benefit of these approaches has proved very popular in the machine learning community, especially for deep models.

An early work to outline a successful strategy for mean field variational Bayes approach for horseshoe priors in linear models is by \citet{neville2014mean}. The key contribution is to identify the approximating densities in closed form and to outline strategies for their numerical evaluations. Variational inference under horseshoe priors has since been successfully used for regularizing deep neural networks \citep{louizos2017bayesian, ghosh17, ghosh2018structured}, where full Bayesian inference appears computationally infeasible at present. \citet{ingraham_bayesian_2016} developed the ``fadeout'' procedure for variational inference under horseshoe priors in undirected graphical models, completely circumventing the need for MCMC. The deepGLM model of \citet{tran2018bayesian} also applies a variational inference procedure for regularizing the weights in a deep neural network whose output is then fed to a glm.

A key idea that facilitates variational Bayes inference in these models is again the fact that if $X^2 \mid a \sim \mathrm{IG}(1/2, 1/a)$ and $a\sim \mathrm{IG} (1/2, 1/k^2)$ then marginally, $X\sim C^{+}(0,k)$. Thus, it is possible to write the hierarchical model for the horseshoe prior density on the weight terms $w_{kl}$ in Equation (\ref{eq:nn}) as a product of normal and inverse gamma densities. The major difficulty in inference in deep models is the fact that the likelihood of the observed data is very hard to characterize analytically, which stems from the use of nonlinear transformations in each layer. To circumvent this difficulty, the variational assumption made by \citet{ghosh17} and \citet{ghosh2018structured} is that a posteriori, the weight terms $w_{kl}$ follow univariate normal distributions and the scale terms $\lambda_{kl}$ and $\tau_l$ follow log-normal distributions. Moreover, the variational posterior factorizes as the product of these univariate posteriors. These works also point out that the formulation of Equation (\ref{eq:non}), which they term the non-centered parameterization, offers critical computational advantages to the statistically equivalent formulation in Equation (\ref{eq:nn}), chiefly because it allows the use of the so-called ``reparameterization trick'' first described by \citet{kingma2013auto} for effectively computing the gradients, which is needed for optimizing the parameters in the variational approximation.

While variational approaches at least try to approximate the true posterior in some sense, in certain situations one may choose to focus simply on point estimates, such as posterior modes. \citet{bhadra2017horseshoe} outlined expectation-maximization and local linear approximation strategies to quickly identify the posterior mode in linear models under a close approximation of the horseshoe prior. These approaches work well when the model is shallow, i.e., when there is only one layer of latent variables. But their uses in deep models appear less appealing.   \citet{li2017graphical}  and \citet{li2019joint} mentioned the possibility of deploying the iterated conditional modes algorithm of \citet{besag1986statistical} to identify the maximum pseudo posterior estimate. This strategy is typically feasible whenever a Gibbs sampler is available and the conditional modes are easy to calculate. Nevertheless, neither paper investigated these strategies either numerically or theoretically. 

\subsection{Available software implementations}
We list some publicly available software implementations for horseshoe and other global-local shrinkage methods for nonlinear and non-Gaussian models in Tables~\ref{tab:hs-imp1} and ~\ref{tab:hs-imp2}, which focus on shallow and deep models respectively. Direct hyperlinks to the code repositories are provided, along with the relevant papers and brief descriptions of the application areas. This complements the list in Table 5 of \citet{bhadra2017lasso}, which focuses on implementations of horseshoe shrinkage in linear Gaussian models. Most available implementations are in high level interpreted languages such as Python, MATLAB or R, although some implementations do make calls to compiled C or C++ shared libraries to ease the computational burden. 

\begin{table}[!h]
  \centering
  \footnotesize{
    \begin{tabular}{|c|c|c|}
    \hline
    Software with hyperlinked github URL & Relevant Papers & Brief Description of Functionality\\
    \hline
    \href{https://github.com/liyf1988/GHS}{\textsc{MATLAB} code: GHS} & \citet{li2017graphical} & Precision matrix estimation in GGMs\\
       \href{https://github.com/liyf1988/HS_GHS}{\textsc{MATLAB} code: HS-GHS} & \citet{li2019joint} & Joint mean-covariance estimation in SUR models\\
           \href{https://github.com/aterenin/GPUHorseshoe}{Scala code using CUDA: GPUHorseshoe}& \citet{terenin2018gpu} & GPU accelerated Gibbs sampling in probit models \\
             \href{https://github.com/donaldRwilliams/GGMprojpred}{\textsc{R} package: GGMprojpred} & \citet{williams2018bayesian} & Projection predictive estimation of GGMs\\
               \href{https://github.com/drkowal/dsp}{\textsc{R} package: dsp} & \citet{kowal2017dynamic} & Dynamic shrinkage processes\\
               \hline
    \end{tabular}%
      \caption{Implementations of horseshoe shrinkage for shallow nonlinear and non-Gaussian models.   \label{tab:hs-imp1}}
    }
\end{table}%

\begin{table}[!h]
  \centering
  \footnotesize{
    \begin{tabular}{|c|c|c|}
    \hline
    Software with hyperlinked github URL & Relevant Papers & Brief Description of Functionality\\
    \hline
             \href{https://github.com/VBayesLab/deepGLM}{\textsc{MATLAB \& R} code: DeepGLM} & \citet{tran2018bayesian} & Fitting DeepGLMs with horseshoe regularization\\
              \href{https://github.com/dtak/hs-bnn-public/}{Python code: HS-BNN} & \citet{ghosh17} & Horseshoe regularization for Bayesian neural nets\\
              \href{https://github.com/zhegan27/dsbn_aistats2015}{MATLAB code: dsbn} & \citet{gan2015learning} & Global-local shrinkage in deep sigmoid belief nets \\
               \href{https://github.com/KarenUllrich/Tutorial_BayesianCompressionForDL}{Python code: Bayesian Compression} & \citet{louizos2017bayesian} & Bayesian compression for deep learning \\
    \hline
    \end{tabular}%
      \caption{Implementations of horseshoe shrinkage for deep nonlinear and non-Gaussian models.  \label{tab:hs-imp2}}
    }
\end{table}%

\section{Conclusions}\label{sec:conc}

Global-local shrinkage approaches have proved vastly successful regularizing models of practical interest in machine learning applications. Most existing works have focused on the linear, Gaussian case. The current paper complements the existing literature by providing a summary of the important theoretical, methodological and computational developments in horseshoe shrinkage \emph{beyond} linear, Gaussian models. Several questions remain open and we end with some possible directions for future investigation.\\
\\
\emph{1. Theory.} We summarized the theoretical optimality properties of global-local priors in linear models in Section~\ref{sec:theory}. While all the methodological papers in Section~\ref{sec:shallow} and \ref{sec:deep} contain varying levels of theoretical support, corresponding notions of optimality in nonlinear and non-Gaussian models are yet to be properly defined and explored. The problem is perhaps more acute for deep and densely connected nonlinear models, such as deep neural networks. It appears intuitive that sparsifying priors such as the horseshoe will lead to theoretically desirable properties in regularizing such models. However, at present this remains a conjecture, despite the empirical success demonstrated by papers listed in Section~\ref{sec:deep}.\\
\\
\emph{2. Computation.} A notable feature in the current development of horseshoe shrinkage for complex or deep models has been the proliferation of variational Bayes and point estimation approaches. Given the extent of computing power at present and the complexity of deep models, this trend is understandable. Yet, Bayesian theory, and indeed Bayesian philosophy, calls for the exploration of the full posterior. Development of computationally scalable fully Bayesian shrinkage methodology for deep models is an open problem at present that deserves more attention. Early empirical works suggest a promising approach in terms for computational scalability of Bayesian inference in deep neural networks is the use of  stochastic gradient MCMC methods \citep[e.g.,][]{yao2019quality}. In terms of software availability, a basic implementation of the horseshoe is available in the TensorFlow platform \citep{abadi2016tensorflow} with a Python interface at: \url{https://www.tensorflow.org/probability/api_docs/python/tfp/distributions/Horseshoe}, which should facilitate integration with large scale machine learning techniques for complex and deep models under a unified framework. At present, most available codes for such models (as reported in Table~\ref{tab:hs-imp2}) appear to be standalone implementations.

\section*{Acknowledgements}
We thank the AE and two anonymous referees for many helpful comments. Bhadra and Polson are supported by Grant No. DMS-1613063 by the US National Science Foundation.

\bibliographystyle{apalike}
\bibliography{hs-review}
\end{document}